# Alternative Strategies for Maximizing the Output of Multi-Junction Photovoltaic Panels

Ze'ev R. Abrams

email: zabrams@gmail.com

*Abstract*— **Multi-junction photovoltaics provide a logical method of increasing the utilization of solar power for a given area. However, their current design and fabrication methods invoke numerous material and cost complexities that limit their potential, particularly for flat panel paradigms. In this paper, three general strategies based on the electrical isolation of the internal sub-layers are described. These strategies involve current or voltage matching the sub-layers by varying of fractional absorption and areal coverage of individual cells within each sub-layer, as well as modifying their combined output using power electronics. A simplified theoretical description of these strategies is provided for pairs of junction materials that allows a more streamlined description of the requirements.**

*Index Terms*— **Solar Energy, Photovoltaic Systems, Multijunction, Tandem Cell.**

## I. INTRODUCTION

**M**AXIMIZING the power output of a photovoltaic (PV) system can be done using either a single semiconducting p-n junction with special modifications [1[, optical concentration [2]-[4], or utilizing a multi-junction (MJ) design [5[-[9] that employs multiple pairs of junction materials. Each has its potential benefits and drawbacks. Exotic processes such as multiple carrier generation, intermediate bands, or down conversion [1] are technologically immature and hold less promise once losses in these materials are considered [10]. Concentrating systems increase the complexity, can reduce the reliability, and are inapplicable to most regions of the world where the amount of direct sunlight is limited (typically only 70% [11]) or attenuated due to cloud coverage or smog. MJ systems can employ optical spectral splitting layers [12], which are technologically immature and expensive, or more simply, utilize the fact that the incoming sunlight is a defined by the photon flux on a *given area*, and utilize a stacked structure [5] that absorbs the polychromatic sunlight in a self-aligned, graded fashion [7]. MJ systems based on a stacked design are therefore the most realistic design for near-future PV systems with maximal output, assuming the cost and material complexities can be solved.

The calculated thermodynamic limits of PV energy conversion have been analyzed for a given absorption area, $A_{Total}$, utilizing an ever increasing number of MJ materials using the detailed-balance [13] theoretical approach, which is an interplay between the incoming and outgoing solar flux, and the current and voltage of each cell [6]. This can be done using the measured AM1.5 terrestrial or AM0 extra-terrestrial spectra, or generalized using the blackbody spectrum of a $T_{Sun} \approx 6000K$ source, and a PV system on Earth, at $T_{amb} = 300K$ [13]. Efficiency analyses for the entire PV *system* are then based on the calculations of the output of a single MJ cell stack, and then linearly interpolated for the entire system. However, most PV systems contain a number of cells in a module, which are then combined electronically in a string of panels. An effective design of such an MJ system should employ this fact to maximize the power output.

In this paper, a set of techniques are described to maximize the output of a PV system constrained to a set area, $A_{Total}$, using a simplified analytical set of formulas so as to focus on the critical aspects of such a system, and to allow an easy assessment of any choice of materials comprising the MJ. Adding an additional sub-layer to a MJ system does not increase the overall efficiency in a linear manner [1], and therefore the choice of bandgap for this additional material is not the only critical factor. Specifically, we describe ways to current and voltage match isolated systems based on absorption variations, areal matching of the sub-cells, and power matching using power electronics. This description requires the perspective of a PV system as consisting of multiple cells per layer, each with its own possible current and voltage. Each of these methods can be used independently, or in combination, to maximize the $\sum I \times V$ product of the entire system.

## II. DESCRIPTION OF A TRADITIONAL TANDEM SYSTEM

The canonical MJ system can be described using a tandem [14[,[15], two-junction device, and any multiplicity of junctions can easily be generalized using these concepts. This terminology will be employed throughout this text, primarily since visually presenting results for more than two junctions is increasingly complex, particularly when trying to generalize for different material parameters. Such a generic tandem configuration is displayed in Fig. 1a, with a top and bottom semiconductor connected in series via a tunnel junction. The current and voltage through this device is current-matched:

$$I = i_T = i_B$$
$$V = \sum v = v_T + v_B \tag{1}$$

This situation is similar to a standard, single material junction, PV panel consisting of multiple cells per panel (e.g., 60 cells), such that the overall power is $P = I \times \sum V$, with the current defined by a single cell in the panel, and the voltage being the cumulative voltage of all the identical cells. It should be noted that this description assumes that the area of each cell, $a_T$ and $a_B$, are equal, so that both the current and current density, $J$, are the same (with $i = a \times j$). This holds true for a standard panel as well, with each sub-cell in the panel being of equal area to match the current. If $J$ is not matched, then the current will be limited by the lowest cell current in the stack.

The properties of a MJ are defined foremost by the constituting bandgaps, $E_T$ and $E_B$ for a tandem MJ, as well as the difference between these bandgaps, i.e. the bandwidth: $\Delta = E_T - E_B$, as shown in Fig. 1b. Using a standard detailed balance approach, it is known that there is an optimal set of bandgaps such that there is a maximal efficiency for the tandem MJ system, as shown in Fig. 1c. This result is for an *electrically isolated* pair of junctions, whereas the result for a constrained, current matched system [15], following (1), is slightly lower [1]. This model assumes that each semiconducting material absorbs all incoming sunlight from its bandgap and above, with a step-function absorption coefficient [13],[16].

The tandem MJ system will result in a pronounced drop in current in the bottom cell, due to the cutoff of absorbed photons above its bandwidth. The exact efficiency of each pair can be calculated using a detailed balance approach, as demonstrated for a Silicon (Si, $E_G = 1.12$ eV) bottom cell in Fig. 2a, which can be viewed as a perpendicular cross-section of Fig. 1c at $E_B = 1.12$ eV. As is shown in Fig. 2a, the bottom cell's efficiency is always lower than what it would have been in isolation (dotted blue line), whereas the top cell is "unaffected", and follows the standard detailed balance calculated curve (dashed red). The bottom cell's efficiency drop is therefore critical when choosing any pair of materials for a tandem system.

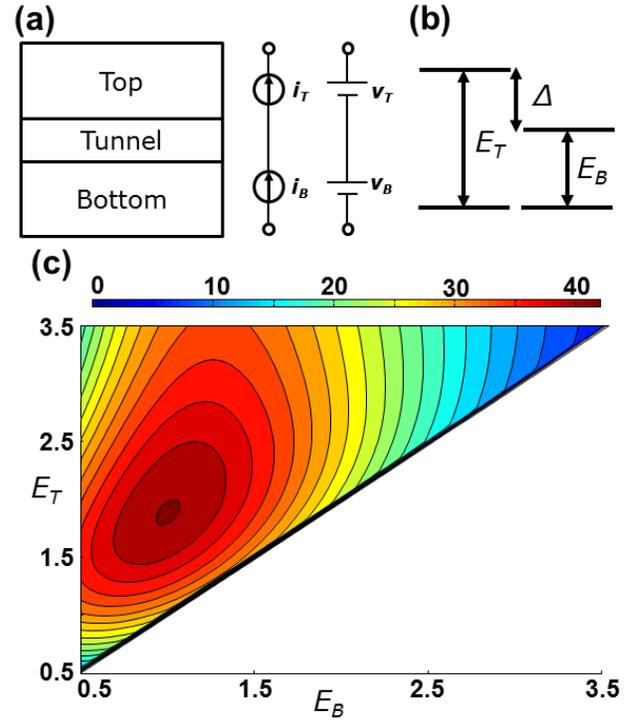

Fig. 1 (a) Depiction of a tunnel-junction tandem device, which can be described as two current sources or two voltage sources. In this device, the current sources are in series. (b) Simplified band diagram of a tandem multi-junction device, depicting the two bandgaps, $E_T$ and $E_B$, as well as the bandwidth window between them, $\Delta$. (c) Detailed Balance calculation for a pair of isolated cells, with a maximum of ~43% using a blackbody approximation of the sunlight.

While the efficiency drop of the bottom cell can be calculated by numerically solving the detailed balance conditions as a function of $\Delta$, we here present a simplified set of formulas that allows a direct assessment of this efficiency drop. The short-circuit current of a single junction material, with bandgap $E_G$, and absorbing sunlight from a blackbody at $T_S = 6000$K is proportionate to the incoming photon flux:

$$i = a \times g \times \int_{E_G}^{\infty} \frac{E^2 dE}{\exp(-E/kT_S) - 1} \tag{2}$$

Where $g$ is a constant to ensure the correct units, and is approximately 345 mA/cm² [with $g = \Omega_S q^4/\hbar^4 c^2$ including the incoming sunlight's étendue ($\Omega_S$) [17], electron charge ($q$), Planck's constant ($h$) and the speed of light ($c$), and with Boltzmann's constant ($k$) in units of eV/K such that $kT_S = 0.517$ eV]. The blackbody approximation is good for analytical formulas, and is an approximation of either the AM1.5 or AM0 spectrum; it is used here to simplify all formulas. Equation (2) can be approximated with good accuracy above bandgaps of 0.5 eV to be:

$$i \cong a \times gkT_S E_G^2 \varepsilon_S \, exp(-E_G / kT_S)$$
$$\equiv a \times j^{reg} \tag{3}$$

where the term $\varepsilon_S$ is due to the approximation for the integral [10],[17], and is given by:

$$\varepsilon_S = 1 + 2kT_S / E_G + 2(kT_S / E_G)^2 \tag{4}$$

In contrast, in a tandem MJ system, the bottom cell only absorbs light within the bandwidth $E_B \rightarrow E_B + \Delta$, and therefore its maximal current is given by:

$$i_B = a_B g \int_{E_B}^{E_B + \Delta} \frac{E^2 dE}{exp(-E / kT_S) - 1}$$
$$i_B \cong a_B gkT_S E_B^2 [\varepsilon_S \, exp(-E_B / kT_S)... \\ - \varepsilon_\Delta \, exp(-(E_B + \Delta) / kT_S)] \tag{5}$$

where the additional correction term, $\varepsilon_\Delta$, is [18],[19]:

$$\varepsilon_\Delta = \varepsilon_S + 2\Delta / E_B + 2\Delta kT_S / E_B^2 + 2(\Delta / E_B)^2 \tag{6}$$

Equation (5) can be further simplified in terms of (3) to be:

$$i \cong a j_B^{reg} \left[ 1 - (\varepsilon_\Delta / \varepsilon_S) exp(-\Delta / kT_S) \right] \tag{7}$$

With $j_B^{reg}$ being the current density in the bottom cell had it been in isolation, i.e. (3) replacing $E_G$ with $E_B$. One can define a general term, $H_\Delta$, that will re-occur in subsequent formulas to be:

$$H_\Delta \equiv (\varepsilon_\Delta / \varepsilon_S) exp(-\Delta / kT_S) \tag{8}$$

such that (7) can be simplified to be:

$$i_B = a_B j_B^{reg} (1 - H_\Delta) \tag{9}$$

The current of the top cell can also be described using this term as:

$$i_T = a_T g \int_{E_B + \Delta}^{\infty} \frac{E^2 dE}{exp(-E / kT_S) - 1}$$
$$= a_T j_B^{reg} H_\Delta \tag{10}$$

where it should be noted that the term $j_B^{reg}$ is a function of the bottom bandgap, $E_B$. The two currents are therefore easily calculated as a function of the original (isolated) current of the bottom cell, $j_B^{reg}$, and the algebraic term $H_\Delta$, using a simpler set of equations. The current $j_B^{reg}$ can be calculated using (2), or using the approximation of (3) [20], with little difference between them. The term $H_\Delta$ essentially represents the current ratio difference between the tandem system current, and the original current that the bottom cell would have produced without the top cell. In addition, it appears in the open-circuit voltage for the bottom cell, assuming zero radiative coupling between cells [18[,[19[,[21[:

$$V_{oc}^B = V_{oc}^{reg} + kT_{amb} ln(1 - H_\Delta) \tag{11}$$

with $kT_{amb}$=25.8 meV. This correction to the open circuit voltage is larger for smaller values of $\Delta$ [19]. The term $V_{oc}^{reg}$ can be calculated using either the direct way [$V_{oc} = kT \times log(I_S/I_o + 1)$] [22], or using well known approximations for the thermodynamic maximal $V_{oc}$ [16],[18], and can be approximated in general for an *ideal* flat cell as [23]:

$$V_{oc}^{reg} = 0.95 E_G - 0.2 + 0.0258 ln(\varepsilon_S) \tag{12}$$

in units of Volts, where an additional loss term of $kT_{amb} ln(\kappa_{loss})$ can be included to incorporate all other losses [24],[25].

The overall efficiency of the system is related to the short-circuit current, open-circuit voltage, and fill-factor (*FF*), with the *FF* being dependent upon the $V_{oc}$ directly [26] as well as being highly dependent upon the material parameters. Cells that aren't produced well typically have low $V_{oc}$s and *FF*s, whereas the ideal *FF* is typically over 80% per junction.

While the $V_{oc}$ is critical for high efficiency MJ systems [18],[19] the predominant factor in defining the efficiency of each sub-cell in the system, and in particular the bottom cell in the tandem structure, is the current. When compared to the precise detailed balance method (for bandgaps above 0.5 eV), the ratio of currents of the bottom cell before and after a top cell is placed above it is nearly equivalent to the calculated efficiency ratio of the bottom cell before, $\eta_B^{reg}$, and after, $\eta_B^{MJ}$. The drop in efficiency between the isolated and stacked bottom cell is therefore, using (9):

$$\eta_B^{drop} = 1 - \frac{\eta_B^{MJ}}{\eta_B^{reg}} \cong 1 - \frac{i_B^{MJ}}{i_B^{reg}}$$
$$\eta_B^{drop} \cong H_\Delta \tag{13}$$

The approximation using $H_\Delta$ is typically 1% more than the precise detailed balance calculated efficiency drop, however, it is much easier to calculate, and simplifies the understanding of what efficiency drop is expected when considering a pair of bandgaps separated by a bandwidth $\Delta$. The efficiency drop for a bottom cell within a tandem MJ system is plotted in Fig. 2b, for various bandwidth separations using (13). Also plotted are the efficiency drops for a Silicon/Cadmium-Telluride system (Si with $E_B$=1.12 eV and CdTe with $E_T$=1.49 eV, $\Delta$=0.37 eV, star), and a Germanium/Cadmium-Telluride system (Ge with $E_B$=0.67 eV, $\Delta$=0.82 eV, diamond). The actual detailed balance calculated efficiency drops of the bottom cells for these pairs are 71.1% and 51.4%, respectively, using a blackbody approximation. Also plotted in Fig. 2b is the exact calculation of the currents (dotted lines), using (2), demonstrating the relative accuracy of using the approximation of the integral given in (3) and then used in (13) [20].

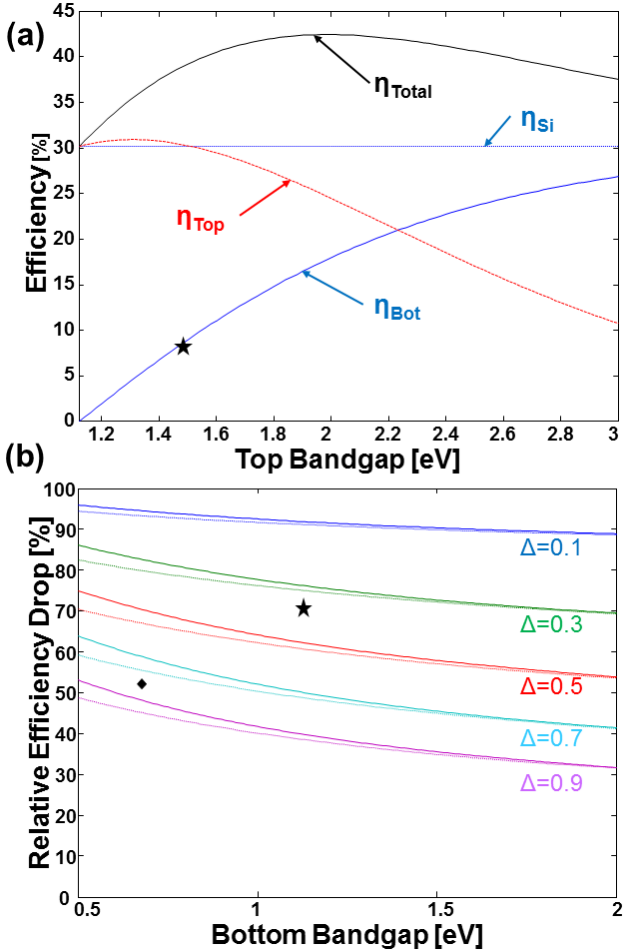

Fig. 2 (a) The efficiency of a Silicon bottom cell ($E_B$=1.12 eV), with varying top cell bandgaps. The total efficiency of the pair ($\eta_{Total}$) is in solid black. The efficiency of the top cell (dashed red), is the Shockley-Queisser detailed balance limit. The bottom cell's efficiency (solid blue) is inversely proportional to the top cell, and is lower than the Si cell in isolation (dotted blue line at ~30%). (b) The drop in efficiency of the bottom cell, compared to its efficiency in isolation, when beneath a top absorber with a bandgap difference of $\Delta$, following Equation (13): $\eta_B^{drop} \sim H_\Delta$. Dashed lines use a numerical calculation of the current, as opposed to the approximation of Equation (3). Plotted also are the efficiency drop of a Si/CdTe system (star) and a Ge/CdTe system (diamond).

Fig. 2 shows that small bandwidths between pairs of bandgaps ($\Delta \rightarrow 0$) results in large efficiency drops in the underlying cell. The overall efficiency of the tandem system is:

$$\eta_{Total} = \eta_B + \eta_T \qquad (14)$$

and therefore, the drop in efficiency of the bottom cell is critical if high cumulative efficiencies are to be reached. Furthermore, this assumes that the top cell is optimized such that its efficiency is as high as possible. This critical point is best illustrated using an example: For a hypothetical Si/CdTe MJ system, the maximal combined efficiency should be roughly 39% (using a blackbody approximation, shown in Fig. 2a). This involves a *reduction* of the Si cell's efficiency from ~30.2% to ~8.7% (as shown in Fig. 2), which can be calculated in the reduction of Maximal Power Point (MPP) current, $I_{MPP}$, due to the smaller absorption bandwidth, from 61.3 mA/cm² to 18.4 mA/cm² (the $V_{MPP}$ is reduced by 0.03 V) [19] when the cells are electronically isolated. The 8.7% efficiency is for an ideal cell with a bandgap of 1.12 eV, and in reality, the best Si cell is limited to 25-29% [27], with current best-of-class cells reaching 25%, leading to an expected efficiency of the bottom sub-cell of (25%/30%)×8.7% = 7.25%, before any other losses caused by the integration of the cells into a module are included.

### III. General Strategy for Matching Sub-Layers

Figure 2 essentially described how adding additional layers to a MJ system do not increase the cumulative efficiency in a straightforward manner. Furthermore, the above analysis of the tandem design demonstrated how the currents and voltages of each sub-cell are generally different. The standard design of a MJ system focuses on individual *cells*, since they are fabricated using a sequential stratified deposition, including the absorbing layers (emitter and base), tunnel junctions, as well as graded layers to match the incompatible lattice constants of the constituting materials [1],[7],[22]. This latter factor is the most critical, and as a result, most MJ systems are built for concentrating PV systems where the cell size can be in the order of mm², and their high fabrication costs can be offset by lower cost optics. The main difficulty then lies with current matching the differing materials, since the overall current of the system is limited by the lowest current in a series-connected circuit consisting of current sources. The surplus current will be lost as resistive heat in the other semiconducting materials, and will therefore reduce the overall efficiency of the MJ system.

If however, we disregard the need for a tunnel junction to help match the currents, different strategies can be employed. These strategies are based upon the idea of electronically isolating each sub-cell, such that a "tandem" MJ system becomes a 4-Terminal (4T) device, instead of a 2-Terminal (2T) one [28]. These 4Ts can then be connected in series, parallel, or neither and then combined using an additional circuit to produce the desired current or voltage output. In addition, this includes the strategy that each set of sub-cells can be of different areas, since the overall current is a function of the sum of areas of each sub-cell. Matching the currents (or voltages), for example, can then be accomplished by varying the relative absorption of the overlying cells, as well as their relative areas, as depicted in Fig. 3.

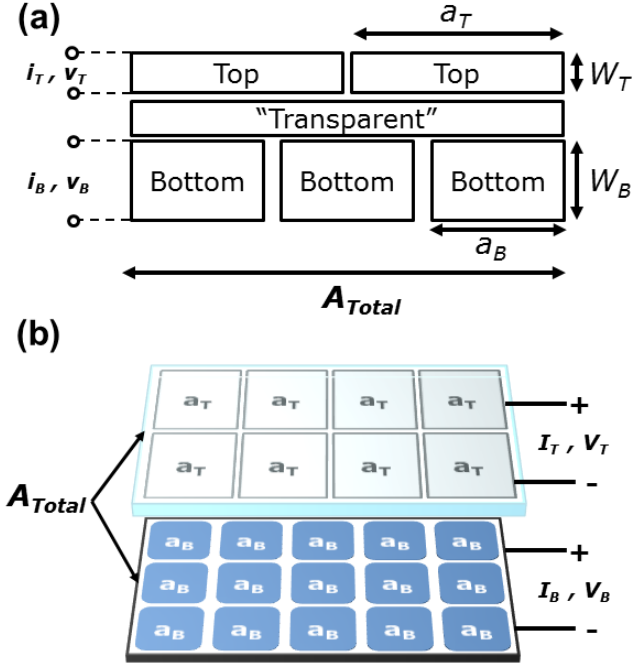

Fig. 3 (a) Side-view schematic of a simplified tandem system, including a top layer of cells that are electronically isolated from a bottom layer of cells via a transparent coupling medium. Each subset of cells has its own area and "thickness", as well as currents and voltages. The overall area of the system is constrained to an area of $A_{Total}$. (b) An isometric view of a theoretical system consisting of two subsets of cells with different areas, thicknesses, currents and voltages, manufactured separately and then combined.

The overall efficiency of a PV system is a function by the surface area of the impinging light, and if we define any system as being a "string" of "modules" that consists of an array of "cells", then each module in the system with an area of $A_{Total}$ must also include $m$ number of cells per internal sub-layer. A flat panel construction is both the most general as well as the best defined for calculating the efficiency [13]. Given a constrained area (either the size of a module frame, or the size of a location to install the PV), the constraint on the sub-cells, if we neglect the spacings between cells and margin areas (which can easily reach 10% of the overall area), is:

$$A_{Total} = \sum_{m_B} a_B = \sum_{m_T} a_T$$

(15)

If we assume that each sub-cell per sub-layer is of the same area, as in Fig. 3b, then the sum can be written as:

$$A_{Total} = m_B a_B = m_T a_T$$

(16)

Neglecting the gap and margin regions needed to isolate sub-cells, this means that the number of sub-cells within a constrained region must be an integer multiple of the individual areas.

Using this set of constraints, we now will show how to current and also voltage match the systems, assuming that they are to be connected in series (or parallel). Subsequently a strategy employing an external power electronics circuit will be presented that encompasses both sets of strategies.

## IV. CURRENT MATCHING WITH VARIED ABSORPTION

The first method to match currents is to vary the amount of absorption in the top material in a tandem MJ system. Here, we will assume equal areas of the sub-cells, $a_B = a_T$. The actual absorption of a material is dependent upon its thickness, $W$, and its absorption co-efficient, $\alpha_{abs}$, or in the effective thickness for the case of a porous substrate or material made of nano/micro-wires [29]. This changes from material to material, particularly if the materials are indirect bandgaps and optical trapping methods are employed [22], which would then be difficult to generalize the approach in an analytical fashion. If however, we view each PV material as acting essentially as a photon-counter, for all photons absorbed above the bandgap, then the exact absorption spectrum of the top layer is not as critical as the overall *number* of photons that then reach the bottom layer. Caveats to this statement are critical in any real system, particularly in regards to the differences in absorption of high energy photons at the surface, as opposed to lower energy photons within the bulk of the material. However, the overall calculation of the expected efficiency of the bottom cell remains essentially the same using a "lumped" absorption coefficient concept for the top cell: $F_{abs}$.

The absorption of each cell can be simplified to first-order as a unity step function [13],[16] with full absorption occurring at the bandgap, as shown in Fig. 4a. The cells are cut-off at high energies due to the lack of photons from the sun at energies above ~5 eV (depicted as $E_{Sun}$). The actual absorption co-efficient of the top cell will vary (depicted as a red line in Fig. 4a), but as far as the bottom cell is concerned, if only (e.g.) $F_{abs}$=80% of the photons are absorbed above $E_T$, then ($1-F_{abs}$)=20% of the photons between $E_T$ and $E_{Sun}$ will reach the bottom cell. The current for the top cell will therefore be:

$$i_T^F = a_T j_B^{reg} F_{abs} H_\Delta$$

(17)

Whereas for the bottom cell, we must re-write the formula for the current to include the extra fraction of photons reaching it as:

$$i_B^F = ga_B \int_{E_B}^{E_B+\Delta} \frac{E^2 dE}{exp(-E/kT_S)-1}$$
$$+ (1-F_{abs})ga_T \int_{E_T}^{\infty} \frac{E^2 dE}{exp(-E/kT_S)-1} \qquad (18)$$

with $E_T = E_B + \Delta$, and which after a simple re-arrangement, becomes:

$$i_B^F = a_B j_B^{reg}(1-F_{abs}H_\Delta) \qquad (19)$$

where again, one can calculate the factor $j_B^{reg}$ using either (2) or (3). In this case, the efficiency drop for the bottom cell simply becomes:

$$\eta_B^{drop,F} \approx F_{abs}H_\Delta \qquad (20)$$

Formula (20) is depicted in Fig. 4b for an arbitrarily chosen value of bandwidth, $\Delta$=0.5 eV. Here, the top black curve ($F_{abs}$=100%) is the same as the middle red curve in Fig. 2b for $\Delta$=0.5 eV, and the linear drop in efficiency follows that of (20), by a factor of $F_{abs}$. While the efficiency drop of the bottom cell will be mitigated by increasing the effective transparency of the top cell, it should be noted that the efficiency of the top cell will drop by the same factor, $F_{abs}$. This situation is therefore preferable if the top cell can lose some relative current in order to raise the lower cell's current.

Absorptions lower than unity can be achieved either by lowering the amount of absorbing material in a given volume, such as with nano/micro-wires [29], [30], or by modifying the thickness of the material using a 1-D Beer-Lambert relation with the top material's absorption coefficient. The lumped model does not include the lower quantum efficiencies and higher absorption of higher energy photons at the front surface of the top cell, due to surface recombination, such as in a CdS-CdTe cell [22].

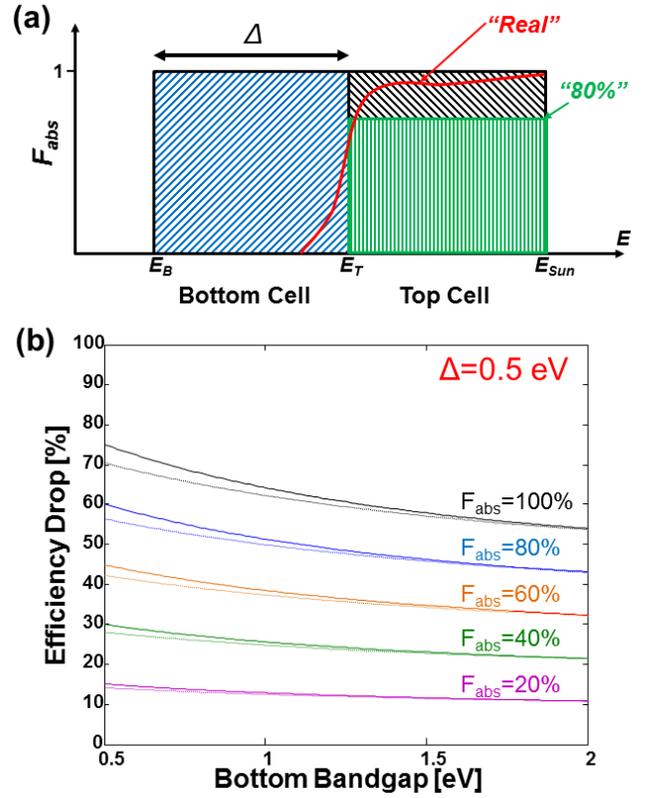

Fig. 4 (a) Simplified lumped model of the absorption of each cell material. The overall absorption of the top cell above its bandgap can be generalized to be $F_{abs}$ the amount of photons that it would have otherwise absorbed. The "real" absorption line is an example of a more realistic version of the absorption coefficient. (b) The drop in efficiency of the bottom cell, compared to its efficiency in isolation, when beneath a top absorber with a bandwidth of $\Delta$=0.5 eV, and assuming a varying lumped model absorption of the top layer. This figure is comparable to Fig. 2b, with the top curve ($F_{abs}$=100%) matching the middle red curve in Fig. 2b.

The difficulty in matching currents between two cells can be seen in Fig. 5a for an example bandwidth difference of $\Delta$=0.5 eV. The black lines represent the current of the top (upper solid line) and bottom (lower dashed line) cells, when $F_{abs}$=100%. As can be seen, these idealized currents never cross each other, with the bottom cell always having lower currents than the top one, regardless of the bottom bandgap (shown here up to 2 eV). In contrast, if we use a lumped model of $F_{abs}$=80% for the top cell, then the extra 20% of above-bandgap photons contribute to a higher current in the bottom cell, as shown in the blue lines in Fig. 5a. In this example, there is a crossing point at approximately 1 eV, such that a pair of semiconductors with $E_B$=1 eV and $E_T$=1.5 eV will have the same currents (again, using the blackbody approximation of the currents).

Using only absorption modulation, one can derive the conditions needed to match currents, assuming that $a_B = a_T$, by equating (17) and (19):

$$a_T j_B^{reg} F_{abs} H_\Delta = a_B j_B^{reg}(1-F_{abs}H_\Delta) \qquad (21)$$

which leads to the condition on $F_{abs}$ that:

$$F_{abs}^{match} = 1/(2H_\Delta)\qquad(22)$$

Equation (22) is an approximation that can allow an "impossible" situation where $F_{abs}$>100%, which essentially means that there is no way to equalize the two currents using this method alone. (22) is depicted in Fig. 5b, where the upper limit is set at 100%. It can be seen that a low bandwidth ($\Delta$=0.1 eV), with two similar bandgap cells, just under 50% of the upper cell must be transmittive (1-$F_{abs}$) in order for the two currents to match. In contrast, for a large bandwidth ($\Delta$=0.9 eV), only the smallest of bandgaps for the bottom cell are possible to enable current matching.

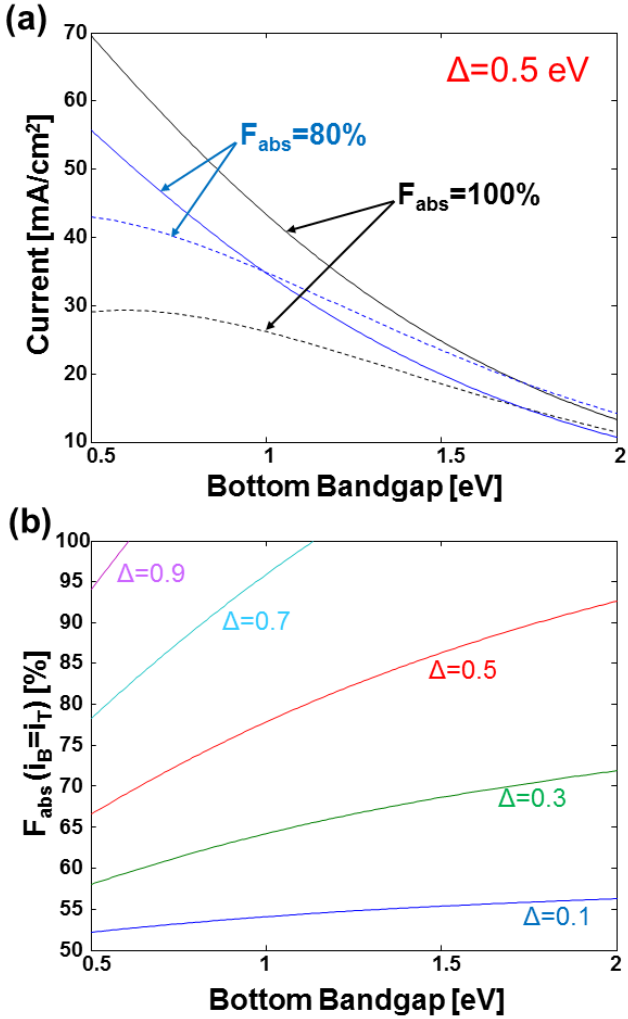

Fig. 5 (a) Current of the top (solid black) and bottom (dashed black) cells with a constant bandwidth difference of 0.5 eV and full absorption of the top cell. No matching of currents can occur in this situation as the lines never cross. For the case where the top cell only absorbs 80% of the incoming photons, (blue lines), there is a crossing point where the currents match at ~1 eV. (b) Mapping of equation (22) in the text, where the current and area of the top and bottom cells are matched. Values above 100% are cut off, as they are physically impossible.

The voltage of the top cell will not drop as quickly as the current, as will be shown in Section IV [equation (27)]. This is easily shown by inspecting the definition of the $V_{oc}$ in relation to the $I_{sc}$ and $I_o$ (dark current): $V_{oc}=kT_{amb}ln(I_{sc}/I_o+1)$. Since the photo-generated current will always be much larger than the dark current, this equation can be written in terms of the fractional top current absorption as:

$$\begin{aligned}V_{oc}^T &\approx kT_{amb}\,ln(F_{abs}I_{sc}\,/\,I_o\,)\\ &= V_{oc}^{T,reg} + 0.0258\,ln(F_{abs})\end{aligned}\qquad(23)$$

Even for low fractional absorptions such as $F_{abs}$=0.1, this would correspond to a drop in voltage of $\Delta V_{oc}$=-60mV.

## V. Current Matching with Different Areas

Allowing the cell areas of each sub-layer to be different, (21) can be used again to match the currents. This can be done with or without the inclusion of thickness/absorption modulation ($F_{abs}$=100% for full absorption). This situation is defined for any system where the following restriction is applied:

$$I_{Total} = I_B = I_T = i_B = i_T\qquad(24)$$

and would otherwise be limited to the lowest current of each set of sub-cells if localized losses such as shading are included. One can re-write (21) as a function of the bottom cell area as:

$$a_T = a_B\,\frac{1 - F_{abs}H_\Delta}{F_{abs}H_\Delta}\qquad(25)$$

Since (25) has too many independent variables to display simultaneously, and since the goal is to demonstrate the results, an example will be chosen: In this case, using the current size of Si cells (as of 2013), assuming that this size will define the sub-cell size of the bottom cells, $a_B$=15.6 cm$^2$, and assuming that the cells are squares (neglecting any corner and gap issues for simplicity), then one can solve (25) for $a_T$ assuming various values of $F_{abs}$. This is plotted in Fig. 6 for full absorption of the top cell (Fig. 6a) and for the previous example of 80% absorption (Fig. 6b). Also plotted on each graph is the area of the bottom cell at 15.6$^2$ cm$^2$ = 243 cm$^2$ for comparison (dashed black line). Depending upon the choice of bottom bandgap and bandwidth, the area of the top cell that ensures current matching of the sub-layers can be above or below the area of the bottom cells.

The choice of $a_B$=15.6 cm$^2$ was completely arbitrary for (25) and Fig. 6, and in general, cells defined using many thin-film technologies can allow a range of cells sizes since their area can be defined post-deposition using scribing lasers. However, Fig. 6 demonstrates the general technique of varying both the area and thickness of each sub-cell, as in Fig. 3, to current match the system.

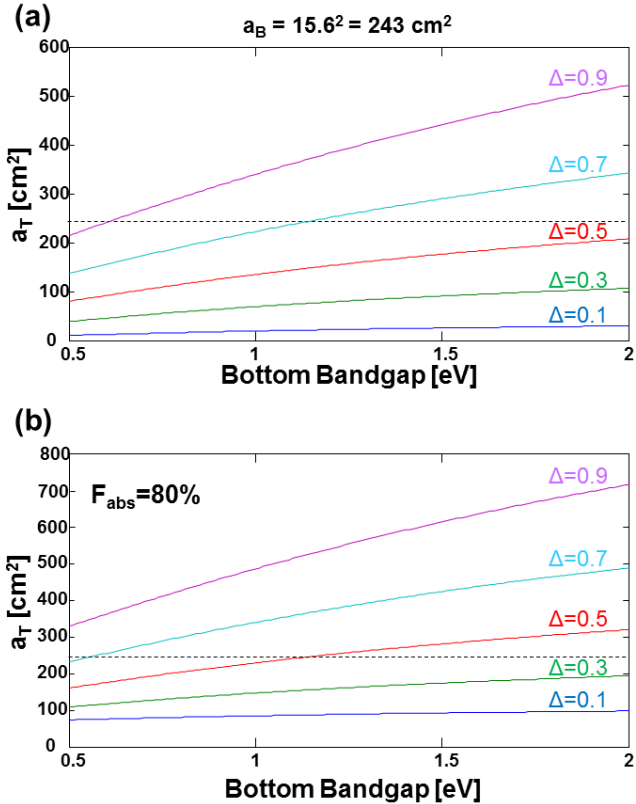

Fig. 6 Current matching of the top cells assuming bottom cells with areas of $a_B$=15.6$^2$ cm$^2$. (a) Assuming full absorption of the top cell, and (b) assuming only 80% absorption of the top cell. The dashed black line is the area of the bottom cell, $a_B$=243 cm$^2$ for comparison. Note the difference in y-axis.

## VI. NUMBER OF SUB-CELLS FOR CURRENT AND VOLTAGE MATCHING

Since the number of sub-cells in each layer must be an integer amount (if the cells are to be connected in series or parallel), then the constraint on the areal matching of (25) can be combined with (16) to derive the number of cells in each layer, given a total system area of $A_{Total}$. In this case, (25) can be re-written as:

$$m_T = \left\lfloor m_B \right\rfloor \frac{F_{abs}H_\Delta}{1 - F_{abs}H_\Delta}$$

(26)

since $m=A_{Total}/a$. In (26), a rounded down estimate of $m_B$ is used to incorporate the integer nature of the number of cells. The number of top cells should also be an integer value, as well as a number that is capable of being closely-packed into the confines of $A_{Total}$, providing an additional constraint.

Using the above example of a bottom cell of area $a_B$=15.6 cm$^2$ and for simplicity, assuming a total module area of $A_{Total}$=1.6m×0.92m =14,720 cm$^2$ (standard module size today), this means that number of (rounded) cells in the bottom layer is: $m_B$=$A_{Total}/a_B$≈ 60 cells. Fig. 7 plots the number of top cells, as a function of these parameters, following (26), for the case of (a) full absorption, and (b) 80% absorption of the top cell. In Fig. 7a, the scenario of closely matched bandgaps, with $\Delta$=0.1 eV, does not appear, since it would involve hundreds of sub-cells, which would be impractical, and lies well outside the graph's axes. In both panels, the comparison should be made with the dashed line, at $m_B$=$m_T$=60, which would signify a matching of the number, and thus, area of the cells. Figs. 6 and 7 are reciprocals of each other, and the matching points whereupon the curves intersect the dashed line are the same in each graph.

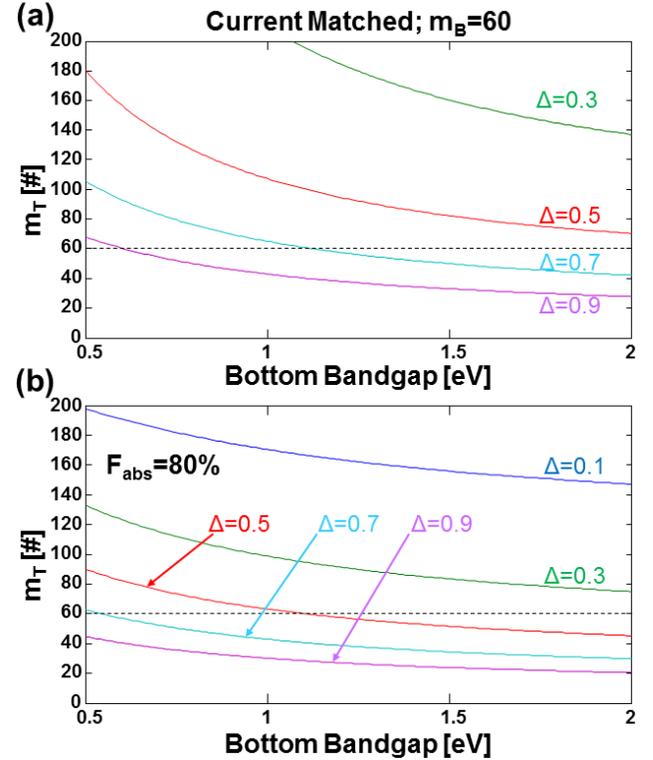

Fig. 7 Current matching of the top cells assuming bottom cells with areas of $a_B$=15.6 cm$^2$, a total area of $A_{Total}$=1.47 m$^2$, resulting in a bottom number of cells of $m_B$=60. (a) Assuming full absorption of the top cell, and (b) assuming only 80% absorption of the top cell. The dashed black line signifies the case where the number of cells is equal in each sub-layer of cells.

One can also calculate the case where the arrays of cells are to be voltage matched to each other. This assumes that the total voltage of the top and bottom sub-layers of the panel of Fig. 3 are to be voltage matched, and assumes that the cells are all connected in series, as in (1). In this case, we can provide a simplified calculation based on the open-circuit voltage (instead of the MPP voltage), using (11) and (12). For this, we can write the $V_{oc}$ of the top cell in comparison to (12) of the bottom cell as:

$$V_{oc}^T = 0.95(E_B + \Delta) - 0.2 + 0.0258\ln(0.5\varepsilon_\Delta)$$

(27)

where the bandgap has been shifted by $\Delta$, the correction term of $\varepsilon_\Delta$ is included in place of the $\varepsilon_S$ term, and the 0.5 factor is included to correspond to the *bifacial* nature of the top cells [31]. Since the sum of the voltages is simply *m* times the $V_{oc}$ of each cell, the voltage matching requirement is:

$$m_T V_{oc}^T = m_B V_{oc}^B \qquad (28)$$

which can also be re-written in terms of the area, $a = A_{Total}/m$:

$$a_T = a_B (V_{oc}^T / V_{oc}^B) \qquad (29)$$

It should be noted that the $V_{oc}$ is nearly independent of the area of the cell, and is a function of the internal properties of the material, temperature and optical conditions [18]. Furthermore, this set of equations did not account for thickness/absorption variations ($F_{abs}$), which can be included, however do not influence the overall $V_{oc}$ by a large amount [19].

Using the same example parameters as above, with $m_B$=60, (29) is plotted in Fig. 8. Note that for this case, the number of top cells is always below the number of bottom cells, regardless of the bandwidth. This is due to the fact that the top cells inherently have a larger $V_{oc}$ due to their larger bandgaps, unless large amounts of losses are included that would degrade the $V_{oc}$.

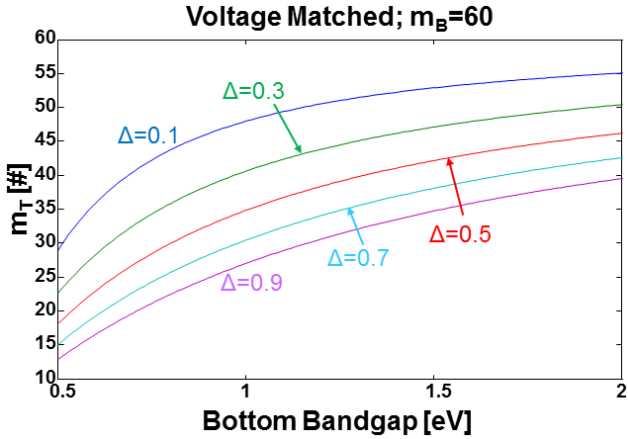

Fig. 8 Voltage matched sets of sub-cells, within a confined area of 1.47 m² and a total number of bottom sub-cells of 60. All values of the number of top sub-cells lie below the number of bottom sub-cells. Voltage matching is here done by matching the open circuit voltages of the cells.

Relying on the voltage of the cells to be equivalent over time has its advantages in that the voltage is independent upon the area, and can remain high even under reduced illumination such as soiling and clouds. However, the voltage of each sub-cell throughout the system is also highly susceptible to variations due to temperature and material defects, and the total voltage will always be limited to the smallest voltage in the circuit. Nevertheless, the change in $V_{oc}$, or, more importantly, $V_{MPP}$, is far less than the change in $I_{sc}$ (or $I_{MPP}$) under varying light conditions, and may therefore be preferable to matching the currents.

Both current and voltage matching examples relied on a constant value for the module area. However, since the number of cells should be an integer amount, the total panel size should ideally be chosen such that the desired number of matching cells can be fully fitted to within the set module area. The size of each sub-cell should also not be made too large in order to limit the series resistance losses that are associated with higher currents and their associated areas.

## VII. POWER ELECTRONIC MATCHING DIFFERENT VOLTAGE SOURCES

The strategies described so far related to the sub-cells within the module. In each of these strategies, the assumption was that the cells must ultimately be connected in series (or parallel) so that they can be combined in a circuit. In particular, one assumption was made that the overall current of the module must be constant throughout the module as in (1). For 2T devices, this can be seen as an advantage in that only two external connections are made; however, the disadvantages were described earlier in terms of the tunnel junction, which considering today's technology, have severely limited the widespread usage of MJs. If, however, one creates an MJ system with *X* number of sub-layers, such that there are 2*X* terminals and connecting wires to the system, and furthermore, if each of these sub-layers is defined by its own set of voltage and current, then the question remains, how to connect them in a circuit at the *system* level?

A strategy to solve this issue is to utilize the advances in Power Electronics (PE), which have enabled the creation of both high power inverters – to convert the Direct Current (DC) producing modules into Alternating Current (AC) sources for the electric grid – as well as DC optimizers that can solve the issue of shading on an array [32]. In essence, a DC optimizer tracks the MPP of each module, or set of modules in an array (string), and varies the *load* proportionately such that the maximal power is extracted from the system, despite mismatches in currents and voltages that can occur due to shading, soiling or other means [33]. The PE is then in charge of recombining these different power sources into a single voltage and current (either DC or AC). One can therefore apply the same general principal to the different current and voltage producing sub-layers of an electronically isolated MJ system.

In Fig. 9, a generalized schematic of such a system is presented, displaying some of the features of this form of integrated "module", combining the concepts mentioned above (areal and thickness variations), as well as an integrated PE system. The schematic can refer to a single module in an array, or to multiple connected modules, such that each sub-layer in each module is connected in series to the equivalent sub-layer in the next module, and not to the sub-layer below it. In this schematic, three distinct sub-layers (whether individual cells, or a layer of such cells, as in Fig. 3b) are shown, each with its own voltage and current. The MPP of each sub-layer of cells can be detected, and the appropriate load applied [33]. In addition, if – for example – the voltages are to be combined into a single voltage, well-known Buck/Boost circuits can be used to either raise or lower the voltages of each separate subsection to match the others [32]. In fact, the final voltage combined from these three elements need not be any one of the constituting voltages. In other words, given three voltages: $V_T$, $V_M$, $V_B$, the voltages can be combined as any multiple of these values, depending upon the desired output. This is accomplished using PE by varying the duty-cycle, $d$, of (e.g.) a Buck/Boost circuit to lower/raise the given input voltage to a desired amount [32]. In particular, for a Buck-Boost circuit, the final output voltage of each sub-layer will then be:

$$V_{out} \propto \frac{d}{1-d} V_{in}$$
(30)

which allows the output voltage of each sub-layer to be above or below the cumulative voltage of all of the internal cells in the sub-layer. It should be noted that some losses will occur in this form of circuit, however, such losses already occur in any inverter and MPP system, and are known as derating factors [34].

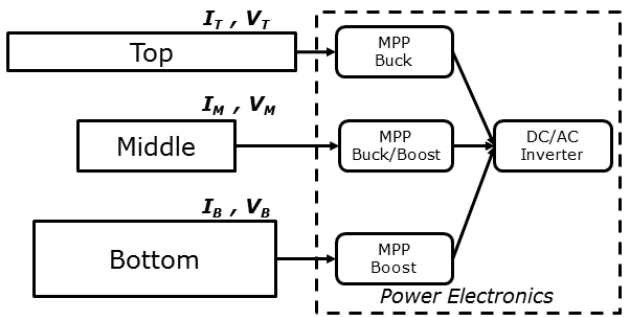

Fig. 9 An example schematic of a system consisting of three sets of sub-cells, each with its own current and voltage, and combined using power electronics, which can include maximum power point trackers and Buck/Boost circuits to match the voltages. The combined output can be then converted by an inverter to create an AC source, or charge a DC battery.

Combining the PE into the design of the currents and voltages of the module *ab initio*, an additional variable can therefore be used to control the output power of the sub-layers: the duty-cycle of each PE sub-system, $d_{match}$. By altering the duty-cycle of each PE sub-system (e.g., each sub-Buck/Boost system), the entire "module" can dynamically react to changes in voltages and currents. This feature allows the complete isolation and optimization of each sub-layer in the array – and can even allow the retrofit attachment of sub-layers on top of existing arrays. Moreover, this feature allows the integrated module to adjust to time-varying changes in the incoming spectrum, which in a tunnel junction MJ results in a decreased power output due to inconsistent current matching. These changes include both shading losses, soiling and changes in the solar spectrum over the course of the day and year. Although this system is no longer completely passive, and relies on an external power source, this distinction is already true of any grid-integrated system employing an inverter, and therefore does not introduce a non-compliant element to the overall system. It does, however, require the complete integration of cell, module, and electronic designs into a unified package. While this design employs more loss incurring components [34], these possible losses would be offset by the advantages this design has in terms of cell mismatches and module soiling.

## VIII. CONCLUSIONS

Maximizing the output from MJs must incorporate variations in the materials, cell designs and module designs. Here were presented a few strategies for these modifications, which can be used independently, or in conjunction, in order to create higher power output systems incorporating existing PV junction technologies. Modifications of the absorption of the top cell are known, however, when taken in conjunction with areal variations, more control over the options to current and/or voltage match the overall system is achievable. More importantly, integrating advanced PE circuits into the design of a MJ system is an important pathway for optimizing the materials and device structures. While other MJ strategies exist, including non-planar geometries [35],[36], the flat panel paradigm without concentration allows the widespread integration of standard module manufacturing and installation into what has become the standard for PV today.

The tandem MJ can be viewed as either adding efficiency to highly efficient back cell, or to a highly efficient front cell. As Fig. 2 demonstrated, the efficiency gain of adding a back cell can be negligible if the bandwidths are not chosen properly. Furthermore, using the standard detailed-balance calculations, it is apparent that unless the back cell is of high efficiency to begin with, the added gain will be negligible. Conversely, if one starts with a high efficiency back cell, and adds a low efficiency top cell, that cell will absorb much of the useful current in an ineffective manner, thereby lowering the gain. Both cells must therefore be of as high efficiency as possible.

Using pairs of today's best controlled materials [27] serves as an example of this:

1. Combining Si and CdTe can utilize today's best technologies; however, creating a (semi-transparent) back contact CdTe cell is difficult. If an 18% efficient CdTe cell is placed on a 23% Si cell, then their combined efficiency would be ~25.2%, which is only slightly higher than the original Si cell, despite adding the manufacturing costs of an entire CdTe system.

2. Ge and CdTe have a better bandwidth match; however, Ge cells are currently not standard for large areas and expensive. Furthermore, The Ge's efficiency will drop by ~50% (from 20.8% to 10.1%), such that it is only useful for a high efficiency, bifacial CdTe cell.

3. A Si/GaAs combination has a small bandwidth ($\Delta$=0.31 eV), less than the CdTe, but making higher efficiency, bifacial, GaAs is possible. A 17% GaAs cell on a 23% Si cell would be required just to break-even in terms of ideal efficiency.

4. A Ge/GaAs cell is better matched in terms of bandwidth ($\Delta$=0.76), like the Ge/CdTe tandem, and has the added benefit of allowing a high efficiency top cell (GaAs can reach >28% efficiency). The Ge would still lose nearly 50% of its efficiency, but the combination could allow the tandem MJ to surpass 30% efficiency. Nevertheless, this combination requires two currently expensive materials.

Tunable materials such as CIGS, InGaAsP [37] and their variants allow much better control of the bandwidths, however, the costs and reliability of such materials is still in the research phase. Likewise, replacing GaSb ($E_G$=0.7 eV) with Ge results in similar cost constraints. Further research must be done on the many technological and cost aspects of these new design strategies to fully realize the potential of non-concentrated, flat panel MJ PVs, as well as the ideal electronics configurations of the associated PECs.